\documentclass[]{interact}

\usepackage{epstopdf}
\usepackage[caption=false]{subfig}
\usepackage{xcolor}
\usepackage[longnamesfirst,sort]{natbib}
\bibpunct[, ]{(}{)}{;}{a}{,}{,}
\renewcommand\bibfont{\fontsize{10}{12}\selectfont}

\begin{document}

\articletype{Preprint}

\title{Digital Infrastructure for Connected and Automated Vehicles}

\author{
\name{Quang-Hung Luu, Thai M. Nguyen, Nan Zheng and Hai L. Vu}
\affil{Faculty of Engineering, Monash University, 24 College Walk, Clayton, Victoria, 3168, Australia}
}

\maketitle

\begin{abstract}
Connected and automated vehicles (CAV) are expected to deliver a much safer, more efficient, and eco-friendlier mobility. Being an indispensable component of the future transportation, their key driving features of CAVs include not only the automated functionality but also the cooperative capability. Despite the CAVs themselves are emerging and active research areas, there is a lack of a comprehensive literature review on the digital infrastructure that enables them. In this paper, we review the requirements and benefits of digital infrastructures for the CAVs including the vehicle built-in, roadside-based, operational and planning infrastructures. We then highlight challenges and opportunities on digital infrastructure research for the CAVs. Our study  sheds lights on seamless integration of digital infrastructure for safe operations of CAVs.
\end{abstract}

\begin{keywords}
digital infrastructure; connected and automated vehicles
\end{keywords}

\section{Introduction}

Connect and automated vehicles (CAVs) carry out the promise to transform the transportation systems and reshape the future of mobility. They can provide a significantly safer, more affordable, environment-friendlier, and less congestion. It has been estimated that the CAVs could significantly lessen the total number of crashes \citep{bertini2016preparing}. On top of that, the integration of automated driving capability with the infrastructure were estimated to reduce the crash-related costs to an amount of hundred of billions US dollar every year \citep{cronin2015connected}. Apart from the immediate social and economic advantages, the deployment of CAVs in synergy with appropriate infrastructure, will significantly transform our way of life, from increasing our mobility to transforming existing infrastructures and urban landscapes 
\citep{taiebat2018review, manivasakan2021infrastructure, golbabaei2021role}.

The relationship between the infrastructure and CAVs is bi-directional. The deployment of CAVs will have a strong impact on the current infrastructures; for instance, when they cooperate in platooning mode, redesigning the road loading. In turn, it is well known that their safe and reliable operation is not possible without the support from a suitable infrastructure, especially when the perception of CAVs has been known to be prone to changes in environment \citep{kopelias2020connected}. Having said that, the deployment of CAVs will likely be incremental and in different developmental stages \citep{farah2018,litman2020autonomous}. Therefore, understanding the impacts of the CAVs on the existing infrastructure as well as addressing key requirements of future infrastructure for their safe operation is a significant research area.

CAVs make use of not only the automated functionality but also the cooperative capability for their safe operations. Digital infrastructure enabling them thus serves as the backbone for their efficient operations. For example, a CAV may use wireless communications to share its location, status and receive updates from roadside infrastructure on the coming traffic conditions, such as information on vehicles, pedestrians and congestion. Such communicated information can be useful in preventing accidents, such as a CAV slowing down or changing lanes upon receiving an alert about a sudden stop by a vehicle ahead. By and large, advanced technologies supporting autonomous navigation including  artificial intelligence (AI) and internet of thing (IoT) rely heavily on this infrastructure. Robust digital infrastructures thus indispensable for maintaining the reliability, efficiency, and most importantly, the safety of CAVs in future traffic conditions.

There have been numerous scientific literature reviews on the operation of CAVs, ranging from its deep learning algorithm \citep{aradi2020survey, chen2021deep, fernandes2021point, gupta2021deep, muhammad2020deep}, localization and planning modules \citep{bresson2017simultaneous, paden2016survey}, testing and safety
\citep{kaur2021survey, rajabli2020software, riedmaier2020survey, tang2023survey}, 
security and privacy \citep{deng2021deep,jahan2019,guo2020,lee2022} to overall operation of CAVs \citep{brown2020,kero2020literature}
shared vehicles
\citep{narayanan2020shared, zhao2020enhanced, guanetti2018control}
.
However, a little attention has been paid to review the relation between the digital infrastructure and the operation of CAVs
\citep{wynand2019guidelines}
. Some recent reviews
\citep{liu2019systematic, manivasakan2021infrastructure, tengilimoglu2023implications}
have highlighted the importance of  infrastructure requirements for the self-driving cars. However, their focus was mainly on the physical infrastructure such as roads and land use, and thus do not consider the digital infrastructure. In other words, the is a significant gap in reviewing digital infrastructure that may support the safe operations of the CAVs. 

In this paper, we review the requirements and benefits of digital infrastructure and the CAVs from the state-of-the-art literature. Our aims are to answer the following two fundermental research questions
\begin{itemize}
\item What are requirements and benefits of digital infrastructure to support the safe operation of CAVs?
\item What are the challenges and opportunities on future CAV-enabled digital infrastructure research?
\end{itemize}

\begin{figure*}[!htbp]
\centerline{\includegraphics[scale=0.35]{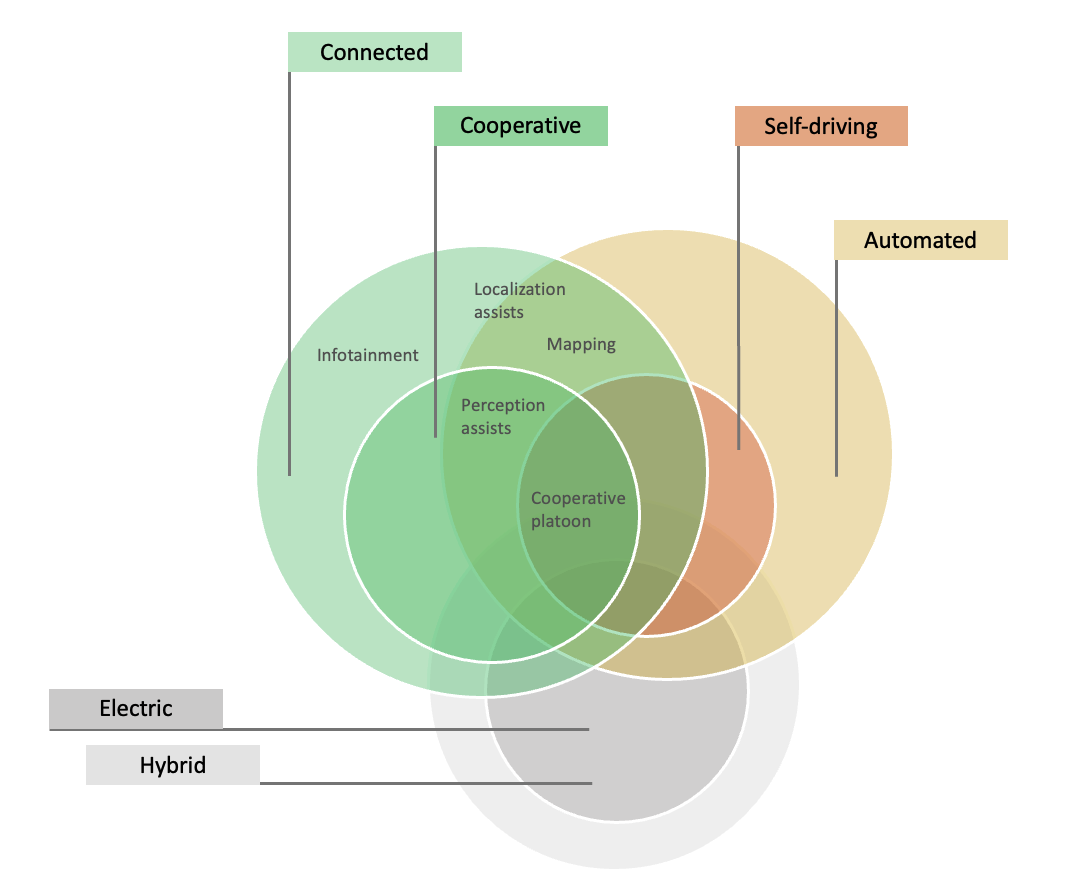}}
\caption{Different types of connected and automated vehicles.}
\label{fig-cav}
\end{figure*}

\begin{figure*}[!htbp]
\centerline{\includegraphics[scale=0.33]{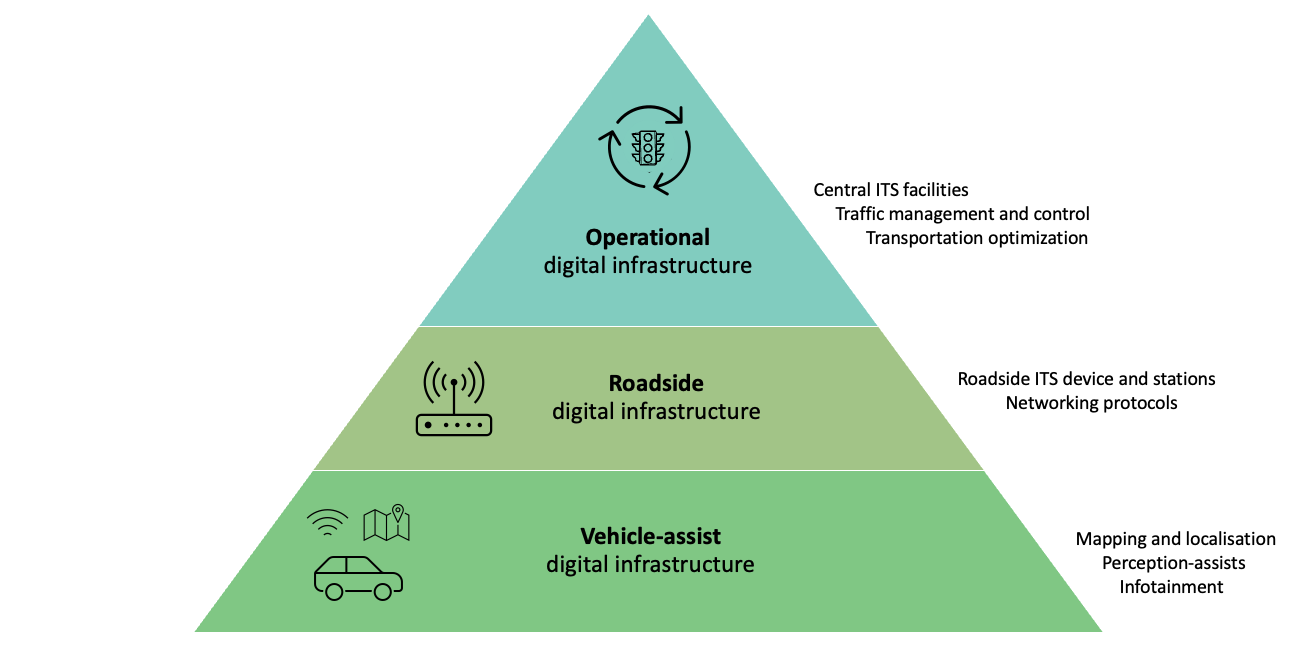}}
\caption{Digital infrastructure to support connected and automated vehicles.}
\label{fig-infras}
\end{figure*}

To address these questions, we consider different types of CAVs, that is, connected-only vehicles and cooperative vehicles  (Figure~\ref{fig-cav}). We also classified the research about digital infrastructure into three types, that is vehicle-assist, roadside and operational digital infrastructure (Figure~\ref{fig-infras}).

\section{Review method}

To perform this literature review, we followed a systematic structured review method \citep{segura2016,kitchenham2004procedures}. Our literature collection covers research works from the beginning of the 21st century on the infrastructure for the CAVs. Our aim was to reveal the key impacts of the CAVs on the existing infrastructure and the requirements of infrastructure for their safe operation. 

\subsection{Search strategy} 

Our review process was summarized in Figure 3. We selected the digital libraries for searching papers based on the following criteria: (i) they were from reputed, and well-established publishers; (ii) they allowed for searching and extracting raw records automatically using application user interface (API); and (iii) abstracts were obtainable with a combination of multiple keywords. Collecting abstracts was important because they gave us better information to filter qualified papers which could not be done by checking their titles alone. 

IEEE’s Xplorer, Elsevier’s ScienceDirect and ACM Digital libraries satisfied all these criteria. To supplement the potential missing studies from other publishers that do not meet those criteria as well as to account for grey publications, we conducted additional searches using Google Scholar by assuming that top results from Google are likely the more relevant grey works. We further expanded the search to 25 first found pages in Google Scholar, each comes with about 20 links. These links were further filtered to exclude the works in three above-mentioned libraries. Some studies mentioned in key papers were also backtracked and collected for inclusion in the review. 

\subsection{Inclusion/exclusion criteria}

We only considered the studies which have been published since 2000, covering a period of 23 years. We combined four sets of keywords (Table~\ref{table:keywords}). The first two categories consist of ``automated'', ``autonomous'', ``self-driving'' or ``connected'' and ``car'' or ``vehicle'' to cover different understanding about the connected and automated vehicles. We use ``digital infrastructure'' in the third keyword. The last keyword is either ``impact'', ``safety'' or ``requirement'', which is one of the main aspects of our study. We combine all four sets of keywords in searching with the ``AND'' operator, where each keyword in a set combined using ``OR'' operator.

\begin{table}[htbt]
\vspace{0.2cm}
\centering 
\caption{Keywords used for literature review}
\vspace{0.2cm}
\begin{center}
\label{table:keywords}
\small
\begin{tabular}{lllll}
\hline
Collection & Set 1 & Set 2 & Set 3 & Set 4 \\
\hline
Keywords & automated & vehicle & digital infrastructure & impact \\
& autonomous & car & & safety \\
& connected & & & requirement \\
& self-driving & & & \\
\hline
\end{tabular}
\end{center}
\end{table}

Note that we ignored specific infrastructure keywords because of two reasons: (i) they gave us more than 10’000 results in IEEE Xplorer alone, which are tedious and impractical to be considered in the review; and (ii) we restricted ourselves to studies that extend the discussion of ``digital infrastructure'' in the general context other than the ones focusing on its specific element. It is possible that our study has not been able to find all papers and materials for such restricted scopes. For example, we excluded the studies which discussed social and environmental aspects of the CAVs, driver and human user interface (HMI), EVs and IoT and edge device in general, as well as physical components of infrastructure. However, we are confident that we bring forward a fair panorama of the state-of-the-art research on the digital infrastructure for CAVs.

\subsection{Publication overview}

The number of publications between 1st Jan 2000 and 1st July 2023 is presented in Figure 4. It was found that within the first ten years of this period, its total number of publications slightly increased, but it remained a small number ($<~10$). We noticed a sharp rise in the number of publications in 2011 when the total number of papers reached 25. From this year onward, a polynomial growth was observed till the end of century. Overall, there was an average of 150 papers having been published each year within the second half of recent decade. The growth was exponential, as compared to the beginning of the century. Those result implied that the trend had taken off within the last 10 years after a decade of gradual development. 

Figure 5 depicted the geographical origin of each study to the affiliation country of its co-authors. We found that the United States led the number of publications with the ratio of 20.9\%, which was almost double that of Germany, the first runner-up with the ratio of 12.2\%. Being ranked in the top three with 10.1\%, China had as twice as many more papers in comparison with other runners-ups, including the United Kingdom, Italy, India, and France. Interestingly, we had not seen many publications from Japan, the home of many well-known car manufacturers including Toyota, Honda, Mazda, and Mitsubishi in this topic, perhaps many documents are either in Japanese or unpublished due to commercial reasons. All three leading countries (the United States, Germany, and China) accounted for almost a half of the publications in the topic of infrastructure for the CAVs.

Figure 6 presents different types of vehicles mentioned in these publications. It was found that the CAVs and the AVs were the most two popular types of names used for vehicles that were relevant to our keyword, accounted for 30.2\% and 28.3\%, respectively. Connected vehicles appeared in about 17.0\% of publications. The wide variety of vehicles mentioned in the study implied a complexity for the infrastructure to support them.

\section{Requirements and impacts of digital infrastructure for CAVs}

\subsection{Vehicle-assist digital infrastructure}

Vehicle-assist digital infrastructure  refers to a digitally-enabled infrastructure that supports and enhances the capabilities of vehicles, particularly in terms of safety, efficiency, and automation. It include the communications between vehicles (V2V), between vehicles and infrastructure (V2I), between vehicles and other entity (V2X) and applications built on top of them to support the capability of the CAVs. It was suggested that the integration of digital infrastructure assisting the vehicle will improve the safety by various indicators. In the United States, it was reckoned that the implementation of early warnings for the road curve to support the CAVs may significantly reduce the crashes by 250 thousand events and 2 thousand fatalities annually; while the integration of warnings for red light violation and pedestrians at crosswalk into digital infrastructure may also reduce 169 thousand crashes and 5 thousand fatalities every year \citep{chang2015estimated}.

High-resolution digital maps of road is one of its elements
\citep{hausler2020map, irannezhad2022minimum}
. While traditional navigation maps such as Google Map, Apple Map, OpenStreetMap being integrated with Geographic Information Systems (GIS) are sufficient for regular connected vehicles, cooperative and self-driving ones require much more sophisticated ones. They includes (i) road shapes and edges with exact positions to keep the CAVs moving within the drivable roads only; (ii) road slope and curvature to assist the CAVs to better control the movement; (iii) road layouts including number of lanes, their locations, sizes, and markings; and (iv) road surface condition including quality measure, maintenance status, gravels, and bumps \citep{hausler2020map, irannezhad2022minimum}. They consist of (ii) speed signs for each lane and road section with detail 3D positions and textual contents; (ii) traffic lights with locations in the transportation network, exact 3D positions in the intersections, specific configuration, and types of traffic lights (i.e., colorings, arrows); and (iii) road warning signs for road conditions such as upcoming merging lanes, giveaway instruction with positions, domains, and schematic information \citep{hausler2020map, irannezhad2022minimum}. CAVs make use of high-resolution and updates of the environment captured in the maps for their precise localizations and appropriate driving decisions in navigating in the real world \citep{huggins2017assessment}. 

Particular attentions should be paid to some important elements of the map. For example, pedestrian crossings to be encoded correctly, especially school zones, shared zones, and high-risk areas \citep{hausler2020map, irannezhad2022minimum}. In addition to the HD maps and the expectation that AVs should also be equipped with higher resolution sensors, more powerful hardware and better deep learning algorithms to cope with complex perceptions and predictions of motorcyclists, bicyclists, pedestrians, and other small-size objects \citep{calvert2020impact}. Enhanced the resolutions of intersections would improve the lane-level vehicle positioning as well as their decisions, and thus safety \citep{liu2013generating}. There were safety concerns with missing detail information in intersections and individual lanes in existing node-based maps \citep{liu2013generating}. 

Obviously AVs at the highest level can take advantage of simultaneous location and mapping (SLAM) technologies to construct the navigation map from combing both physical information and their built-in HD maps, and thus require complex algorithms and expensive hardware. Connected-only vehicles alleviate this demand, requiring information from cellular and roadside infrastructure to facilitate their localization. If there is a gap between digital map and the real-world, especially when the conditions of physical road are deteriorated or altered, an integration of advanced algorithms can be implemented to help the CAVs responding with them \citep{so2023analysis}. Further more, connected vehicles may further use trajectories of other vehicles from the crowd such real-time drivable road conditions reported by other ones in real-time for their decisions \citep{masino2017learning}. 
To provide better real-time road information, sensors could be built into encoded asphalt materials of the pavements as part of smart solutions to inform the CAVs about the road conditions for better driving maneuvers as well as updating the road authorities on their quality in real time \citep{gopalakrishna2021impacts}. 

CAVs should be equipped with at least two DSRC radios and a GPS receiver, which are helpful for providing their speeding and trajectory to enable communication capability \citep{harding2014vehicle}. In addition, it was suggested that a safe integration should consist of an inertial measurement unit to measure the acceleration of CAVs and a human-machine interface (HMI) that informs drivers or passengers on any arising issues \citep{harding2014vehicle}. Using appropriate communication and networking technologies, the CAV can collaborate with others during transportation. There are two types of messages the be part of such V2V communications: the safety message (which can be read by all vehicles) and the certificate exchange message (which is to assure the trustworthiness of the source). For cooperative adaptive cruise control (ACC), the speeding and positioning information of leading and surrounding CAVs and supporting objects transmitted through the DSRC as well as the improved algorithms will be able to extend the existing ACC system to drive the CAVs cooperatively \citep{wang2018review, zheng2016development}. It is also used for warning systems including cooperative forward collision warning (CFCW), intersection collision avoidance warning, highway rail-intersection warning, rollover warning, emergency warning system, as well as approaching emergency vehicle warning \citep{iranmanesh2016robustness}. 

Availability and reliability are key aspects of non-functionality. For safe operations, certain map information should be accessible online, especially for connected-only vehicles. In working under a car park or a tunnel where 5G/6G and GNSS signals are weak, it was proposed that the CAVs can be equipped the capability to localize using SLAM or non-cellular navigation \citep{adegoke2019infrastructure}. Agreements should be made on who provide the road information, which components of map should be stored offline in the CAVs, and which specifications are for real-time over-the-air transmissions. In addition, the infrastructure is also prone to cybersecurity, as many studies pointed out that the digitised transport infrastructure can be manipulated by evasion and poisoning techniques where the map information can be modified in a harmful way, either with the map data stored in the servers or during the V2I communication process \citep{minh2021,deng2021deep}. In the scenario when CAVs collaborate with each other in a cooperative driving, sensor data shared V2V and V2X communications can also be attacked to introduce misinformation on the map and digitalized road conditions \citep{chowdhury2020attacks, noei2016decision}.

\subsection{Roadside digital infrastructure}

Roadside digital infrastructure refers to digital and electronic systems installed along roads to support both conventional and autonomous vehicles. It usually serves a variety of purposes, from enabling vehicle-to-infrastructure (V2I) communication, managing traffic, disseminating real-time transportation information to data gathering. They include sensors to measure traffic and weather, digital signage to display real-time traffic condition; speeding limits and hazard warnings; connected traffic lights that can communicate directly with autonomous vehicles; surveillance cameras for monitoring traffic conditions and events, and wire/wireless networking equipment to facilitate V2I, V2V and V2X communications. The infrastructure should be served as a connection point for data exchange between the CAVs and the security credentials management system (SCMS) to receive security information, update driving behaviors; and should also be allowed for broadcasting and communicating data with other elements and among other V2I applications. The addition of V2I connectivity to AVs may help significantly improve the traffic flow, especially at intersections. Using a car-following model, \cite{ci2019v2i} found that the CAVs can obtain traffic condition in prior to signalized intersections to optimize the movements, and hence enhance the traffic throughput. In general, it was reported that the combinations of individual V2I applications could bring benefits in optimizing the signal timing, reducing the travel time and overall system delay in different aspects: (i) the optimization of signal controls could reduce the travel time by 27\%; (ii) the adoption of the incident guidance for emergency responses may cut off the travel time for emergency vehicles by 23\% and their number of stops by 15\%; and (iii) the cooperation of CAVs by sharing their cooperative adaptive cruise control and adopting a shared speed harmonization may reduce the travel time in highways by 42\% \citep{chang2015estimated}. Cooperative V2I warning systems could help maintain speed before entering the fog area for safe traffic operations \citep{zheng2016development}.

New generations of intelligent transport system (ITS) devices should be developed and installed along the roads to keep track of operations of the CAVs, including the ones to monitor, control and optimize the traffic flow and incident should, and the equipment to monitor road and environments conditions. Many collaborative/cooperative ITS systems have been developed, combining different wireless technologies, to improve safety, transport efficiency and comfortability
\citep{gueriau2016assess, tahir2022implementation}
. They are levels of cooperation between CAVs and infrastructure: traffic adapted-ADAS, V2I connection and V2X communications (Guériau et al., 2016) \citep{gueriau2016assess}. Information on road works, hazardous events, traffic incidents and other issues should be publicly available in the ITS in advance and/or in real-time \citep{irannezhad2022minimum}. A minimal installation of road-side equipment (RSE) related to traffic lights and stop signs at intersections are expected \citep{harding2014vehicle}. In highway, an integration with the CAVs should consist of new deployment or significant upgrade of physical road-side devices for the capability of direct V2I communications, including traffic detectors, environmental sensors, traffic lights and signals, ramp metering system, grade-cross-warning systems, travelers information stations (TIS) (a.k.a., highway advisory radio (HAR) in the United States)
\citep{gopalakrishna2021impacts, manivasakan2021infrastructure}
. Transport authorities may encourage motorcyclists, and perhaps also cyclists, to equip their vehicles with anonymous positioning devices that can be communicated with the infrastructure for safer trips as shown in a pilot study in an urban region of the Netherlands that can enhance safety significantly \citep{calvert2020impact}. 

The advance of VANeT applications bring various benefits to the drivability of the CAVs and the infrastructure. \cite{hu2017link} compared the performance of two VANeT technologies, that is the LTE V2X and the DSRC. Their simulation results showed that both the LTE V2X technology and DSRC can provide low block error ratio (BLER) signals. The LTE V2X outcomes are slightly better than the DSRC by means of lower signal-to-noise ratio (SNR), smaller receiving power, and larger coverage area. The applications of V2X communications with the support of 5G technologies will improve throughputs, reduce latencies, and enhance reliability for a large number of concurrent users \citep{giannone2020orchestrating}. The integration of a moderation system such as Connected Vehicle Service Orchestrator (CVSO) may enable the robust integration of multiple services as ADAS and HD Video Streaming with the support of recent technologies such as Multi-access Edge Computing (MEC) \citep{giannone2020orchestrating}. CAVs can also make use of Vehicular Ad Hoc Networks (VANET) to share media contents across vehicles.  

Although the optimization of CAVs may help improve the traffic in general, communication issues may negatively delay it. \cite{so2015development} pointed out that delays in V2V/V2I reduce the effectiveness of driver warnings by 3-13\%; and in turn, the driver warnings will decrease the conflictions by 27-42\%. The interference of radio frequency on the CAVs and the GPS signals may cause some certain risks to their operations \citep{kar2014detection}. \cite{fernandes2012platooning} demonstrated that communication issues can be addressed if the information from both the platoon's leader and the following CAVs can be employed effectively by adjusting the network load over the CAVs. \cite{hassan2011performance} attempted to improve the performance of the medium-access control (MAC) protocols with DSRC, showing that the revised protocol may significantly increase the traffic density by two times and enhance the signals’ packet delivery ratio by up to 10\%. \cite{abdelhalim2019} proposed an efficient algorithm that helps increase the efficiency and reliability of communications for the CAVs. 

Standards and protocols are backbone that influence the deployed digital infrastructure. They are the rules and guidelines that ensure compatibility and interoperability between different systems and technologies, ensuring that all elements of the digital infrastructure can work together seamlessly. In particular, communication technologies to support the implementation cooperative services in diverse scenarios are critical for the successful deployment of the CAVs, especially the CVs \citep{sepulcre2018context}. The communications and networks infrastructure consists of two aspects, the communication protocols and infrastructure to enable the CAVs, and their applications. First, fast and secure wireless technologies are required for the safe, efficient, and collaborative communications of CAVs with the infrastructure (V2I), other CAVs (V2V) or objects (V2X) including cellular communications, dedicated short-range communications (DSRC), conventional 802.11 wi-Fi, Bluetooth, satellite communications, and radio broadcasting services
\citep{dey2016vehicle, huggins2017assessment}
. Offloading the Internet traffic to DSRC communications can keep the cellular infrastructure from being overwhelmed and inefficient. Cellular V2X (C-V2X) technology will be used, which adopts the cellular protocols as the standards for direct communications between the CAVs and infrastructure (e.g., roadside transmitters) as well as surrounding objects (such as surrounding vehicles, motorbikes, road workers) \citep{baron2016offloading}. \citet{chang2015estimated} suggested that the DSRC can be used for various purposes including mobility application to calculate terminal wait times of vehicles; environmental applications to communicate the locations, speeds, accelerations of CAVs; safety applications to enable crash alerts and warning in real-time; road weather applications to use the CAV sensors as the input data for the operations. Traditional technologies such as radio frequency identification (RFID) technology to help localize the CAVs as part of RFID systems on roads \citep{qin2021collision}. 

With the rollout of emerging of 5G cellular networks, the V2V, V2I and V2X communications are bringing more benefits to the CAVs. With on-going development of 6G technologies, the direct speedy communication and artificial intelligence (AI) will significantly revolutionize the operations of CAVs  \citep{he20206g}. Massive numbers of CAVs as well as their Internet-of-thing (IoT) devices can communicate in grant-free and non-orthogonal multiple access (NOMA) mode. CAVs may enjoy higher data rates with 6G network by taking advantage of mmWave and THz frequency bands in the cell-free mode where access points are deployed along the roads and collaborate through a central processing station to serve the CAVs operating in the covered area. In the cooperative context awareness, intensive AI-based perception tasks will be offloaded from multiple CAVs to mobile edge computing (MEC) nodes using 6G communications such as forwarding sensor to the MEC, and hence the main driving tasks will be reduced significantly. Remote driving functions, edge driving task and, digital twins of CAVs can be supported with 6G. In turn, the CAVs may help extend the coverage 6G networks to through a platform that connects CAVs and unmanned autonomous vehicles (UAVs) as mobile base stations (CBS), especially at night where transport demands are low, in a self-organizing and cooperative mode. 

\subsection{Operational digital infrastructure}

Operational digital infrastructure for transportation  refers to the range of digital tools, systems, and technologies that support the smooth and safe coordination of transportation systems. Traffic management systems are a key component of them which helps collect, analyze, and deliver data about traffic conditions, such as congestion, construction, accidents to support CAVs in making better decisions. They may also include communication systems and cloud-based infrastructure to support central stations to store, process, analyse, optimize and predict traffic. At a higher level, they can be integrated with digital twin technologies in creating a virtual model of the city's infrastructure to predict and optimize traffic flows, simulate various scenarios, and develop more advanced navigation algorithms for autonomous vehicles. 

Another important aspect of traffic control and optimization is about handling the traffic jam, congestions, and accidents
\citep{hendrickson2014connected, puylaert2018mobility, cui2021scalable}
. \cite{hendrickson2014connected} reported that the adoption of CAVs may eliminate about 40\% of congestions during peak hours. A digital and convenient reporting system is required so that road users can report infrastructure issues that may affect the operation of CAVs, such as degradation of dynamic message signs, vegetations in traffic signs, low quality of markings \citep{irannezhad2022minimum}. If traffic flows are efficiently coordinated throughout roadway network, the road capacity can be increased by a similar percentage. Higher efficiencies are observed round bottleneck areas such as tunnels, construction areas, intersections, where the congestions happen more frequent. It was suggested that the crashed-related accidents in highways can be reduced by with appropriate V2I and V2V-enabled traffic management applications, even for the CV or the CAVs at a low level of automation \citep{chang2015estimated,rahman2019safety}. \cite{guo2021drl} suggested that the coordination of the CAVs can be optimized to maximize the green lights and reduce the start-up lost time for increasing the traffic throughputs just via the applications of deep reinforcement learning at the intersections. Since most CAVs are electric vehicles, the traffic flows can be impacted significantly by their battery capacity, charging speed, and vehicle sources with existing infrastructure \citep{miao2019autonomous}. On the downside, \cite{overtoom2020assessing} pointed out that the CAVs may use curbside to drop off passengers, which may lead to local traffic congestion.

The improvement in traffic flow was pointed to be dependent on the penetration rate of the CAVs \citep{mcconky2019don}. If only the lane management is optimized, there will be an increase of 3.6\% in throughputs with the penetration rate of 20\%, while the enhancement is 14.2\% in the scenario when all CAVs are deployed \citep{khattak2022active}. \cite{hoang2022optimal} indicated that the operation of CAVs can provide a better traffic flow even at low penetration rate; and in general, reduce the total travel time of traffic users. In highways, \cite{shi2016autonomous} estimated that significant improvements are noticed only when either AVs share reaches 7\% or the ratio of CVs is 3\%; whilst the impact is only marginal at smaller penetration rates.

One of the notable benefits of operational digital infrastructure is that it can be taken advantage to coordinate the CAVs in the platooning mode. In this mode, the speed differences between the CAVs are minimized while the distance and the stability of traffic flow are maintained. Such modes expected to have suitable communication technologies and network infrastructures, as well as advanced algorithm to moderate the cooperative driving. \cite{jia2016platoon} developed a platoon-based cooperative driving model using realistic V2V connections. In doing so, they proposed novel algorithms for controlling multi-platoons cooperative driving. \cite{jia2018multiclass} further developed a microscopic model for platoons which include the interactions between the conventional human drivers and CAVs and examined their stability condition with respect to the communication delay as well as the penetration rate of CAVs. \citet{sepulcre2016} and \citet{sepulcre2018context} suggested that the communication ranges of fully automated vehicles are better than the conventional ones in the platooning mode; however, it is not the case with the CAV at level 3.

Autonomous car sharing is also one of promising and emerging applications of the CAVs where multiple road users can share the vehicles. \cite{glotz2016electrification} estimated that European cities could reduce the parking needs of 600,000 cars with the car-sharing services. In an analysis on the potential benefits of the CAVs in Adelaide (South Australia), \cite{kellett2019might} estimated the impact of adopting CAVs in the long-term scenarios where all vehicles are replaced by the CAVs with appropriate car sharing and traffic coordination; and found that the city may only need as many as 18\% of the total number of the vehicles as of today, even during the existing morning and evening peaks. If there no agreement is made on the ride sharing, the CAVs still help reduce the on-road vehicles by 27\%, that is, only about two thirds of current fleet could be enough to serve the traffic during peak hours. In the transition scenario in which only two thirds of the vehicles are CAVs and the attitudes to vehicle sharing persist, the city may just need about 82\% the vehicles to maintain the same transportation demands. 

Decentralized vehicular networks have been proposed in different forms, allowing the CAVs with information obtained locally. \cite{gerla2006vehicular} proposed a combination of both vehicular and internet in a so-called Vehicular Grid in order to allow the CAVs to operate seamlessly in emergency operations (e.g., natural disaster, terrorist attack). At intersection, \cite{khoury2017practical} proposed approach to make of only local access information at isolated intersections to make decisions with the application for safe turning.
\cite{raza2018social} suggested to integrated social behaviors into V2X communications to enhance safety. \cite{vanmiddlesworth2008replacing} proposed that pure communications between the CAVs can eliminated the use of infrastructure in travelling through intersections. Interestingly, the CAVs can even use wireless technologies to transfer power across vehicles \citep{abdolmaleki2019}.

From the mobility perspective, An estimation on how the CAV deployment affects the traffic network and the flow provides a critical input for the operational infrastructure. With the involvement of CAVs, the traffic network can be optimized to save the travel time
\citep{chen2017optimal, hoang2022optimal}
. \cite{chen2017optimal} developed a model to optimize the traffic flows with the CAVs and found that the travel costs will be reduced substantially if the CAVs are well operated. They showed that with a simple network of 81 nodes and 288 links where even only a certain area (not entire network) is dedicated to the CAVs, the time decreased 2.36 times in the AV-specific areas, whilst the region outside these areas also experiences a deducted percentage of 20\%. Another simulation study of \cite{szimba2020assessing} also reached a similar conclusion with reduction of 27\% of travel time for the automation at level 5, and 20\% for the CAVs operating at level 4. 

An connected Advanced Driver Assistance Systems (ADAS) can significantly improve the traffic flow when the penetration rate is medium or high
\citep{dey2016vehicle, wei2017dynamic}
. Equipping the CAVs with the ability to send Cooperative Awareness Message (CAM) and to receive traffic signals from Intelligent Traffic Signals (iTS) will save travel times (5\% on average) at intersections \citep{calvert2020impact}. The platoon of CAVs should be equipped with an ability to resolve the contentions, especially at an intersection \citep{lam2013cooperative}.

Interestingly, CAVs themselves are able to be used for controlling and moderating the traffic, the idea may bring benefits to all traffic participants. The information on planned actions of preceding vehicles is greatly helpful to improve the traffic stability \citep{zhang2021dynamics}. The traffic flow in car-following models can be stabilized linearly in the idealized case where there is no model system noise \citep{cui2017stabilizing}. It was also pointed out that a single CAV can help moderate a platoon of an estimated range of 20 to 40 conventional vehicles controlled by human drivers.

\section{Challenges and opportunities in research on digital infrastructure for CAVs}

\subsection{Governmental guidelines and laws on CAV-enable digital infrastructure being either non-existent or unready}

The first critical gap is the non-existing governmental guidelines or laws on CAV-enable infrastructure in most countries around the world. They are emerging now, but all are under exploratory and different across a wide range of practices. This fact subsequently leads to complications in the supporting infrastructure and its related research.

Each country or region has a different principle, guideline, and policy on how to accommodate CAVs. The U.S. Government has outlines principles for developing the CAVs \citep{usdop2021} with three core interests: (i) protect users and communities by prioritizing safety, emphasizing security and cybersecurity, ensuring privacy and data security, enhancing mobility and accessibility; (ii) promote efficient markets by remaining technology neutral and protecting American innovation and creativity; and (iii) modernize regulations by facilitating coordinated efforts, promoting consistent standards and policies, ensuring a consistent federal approach, and improving transportation system-level effects. In line with that, U.S. Department of Transportation has developed various technical programs that support the transitions of infrastructure toward a safer, more efficient, and eco-friendly CAV-enabled transportation \citep{chang2015}. In the dynamic mobility applications (DMA) program, its aim is to collect and disseminate the data from multiple sources from connected vehicles, infrastructure as well as travelers to improve efficiency, mobility, and safety. In the environment real-time information synthesis (AERIS) program, its application is to encourage the development of technologies for connected vehicles, infrastructure and their related applications that reduce the fuel use and emission, such as eco-signal intersection, eco-lanes, eco-traveler information or eco-integrated corridor management. The V2I safety program is to support research and development of technologies that prevent crash through making use of low-latency DSRC applications with applications such as traffic control device, violation warning, stop sign gap assist, pedestrian-enabled signaled crosswalk, curve-speed warning, spot weather impact warning, work-zone warning. In road weather management. Another aspect is to understand the impacts of weather on road infrastructure and ways to mitigate them. They include weather data environment (WxDE), vehicle data translator (VDT) and integrated modelling, variable speed limits (VSL) and traveler reporting for weather responsive traffic management, enhanced maintenance decision support systems (EMDSS), motorists’ advisories and warnings (MAW). 

In other country, such as Australia, the deployment of CAVs is not fully supported by the law, and its existing infrastructure has been designed for conventional vehicles with human drivers \citep{ntc2022}. In the past, inconsistencies have been observed in Australia infrastructure (e.g., three different gauges still existed in rail systems) which caused many efficient issues for travels and transport, which are not solvable by the CAVs \citep{bolsin2018designing}. A regulatory framework has been developed \citep{ntc2022} in line with the Automated Vehicle Safety Law being proposed earlier \citep{ntc2021} through 11 principles: national consistency, alignment with international standards, deployment support, effectiveness with a focus on safety, flexibility, consistent rules, adaptation with changes, clarity on responsibilities and liabilities and efficiency. The framework and the law will define the infrastructure and how the CAVs will be safely deployed and accessed the roads in their interactions with the infrastructure and involved traffics. For instance, CAVs are required to interact with roadside enforcement safely and predictably; and as a result, roadside enforcement protocols are expected to be developed by both national and state-level governments for safe practices \citep{ntc2021}.

While the federal and state governments may have some generic guide and policies to promote the CAVs, the successful integration of the CAVs into the infrastructure is ultimately determined by the local governments \citep{botello2019planning}. Across sectors, the choice of communication standard also affects the optimizations of CAVs. In a simulation-based study, \cite{basaure2021spectrum} pointed out that the CAVs making use of local licensing of 5G spectrum may achieve better performance than the ones using alternative traditional spectrum assignment which is assumed to lower implementation costs while promoting competition.

\subsection{Standardization of CAV-enabled digital infrastructure being nonuniform across countries, regions, and sectors}

In relation to Section 6.1, different understandings, expectations, and technologies complicate the standardization for the CAVs and their related infrastructure. It is of a great challenge in unifying pavement markings alone. The progress and aspects of adoption of modified specifications to the pavement markings varies across different states, which are observed in the United States  \citep{gopalakrishna2021impacts}, Australia \citep{irannezhad2022minimum} and other places \citep{liu2019systematic}. There are also various debates about the quality level of markings and which patterns should be uniformly used. Topics of quality concerns include the durability, the contrast levels, the colorfastness (i.e., fade-resistance), the visibility under wet and glare conditions and the compatibility to the vision of LiDAR and other sensory systems \citep{irannezhad2022minimum}. There is a missing guideline on the uniformity reflected in the traffic signals, such as the existence and the flashing or permanency of the arrows in traffic lights \citep{gopalakrishna2021impacts}. In addition, agreements have not been reached on various domains, such as a uniform chevron marking for gore areas, acceptable joint contrast patterns and the tightening of delineation of special lanes such as toll, bike, or high-occupancy vehicle lanes. 

At the initial stage, Austroads proposed minimum infrastructure requirements and further suggest that the gap in physical infrastructure could be filled by digital infrastructures if a minimum upgrade is preferred \citep{irannezhad2022minimum}. For a cost-efficient roadmap toward integrating CAVs into the infrastructure, the changes are classified into four groups for all infrastructure elements, namely basic upgrades (Class 1), basic new infrastructures (Class 2), AV-specific upgrades (Class 3) and AV-specific new infrastructure (Class 4). Specifically, those changes are implemented by three main phases. For the short-term from now to 2030, changes associated with Class 1 and Class 2 should be implemented first. In general, improvements include upgrades in pavement markings and work zone signs. Changes should also include ramp marking and signages, safe stopping areas and interchange hubs/points in motorways and urban highways along with pavement contrast ratio, one-off signs, and graded intersections in rural highways. In urban areas, one would also consider upgrades in traffic signals, school zone signs, pickup/drop-off points for CAVs and related footpath facilities. For the medium-term from 2030 to 2040, CAV-specific upgrades are expected in bridges and pavement designs in all roads. In urban areas, improvements are needed for transit or semi-transit dedicated lanes for CAVs. For the long-term from 2040 onward, digital infrastructure should be built across Australia. In urban roads, mobility hubs/points for CAVs and remote facilities will be of high interest, subject to the demand. In highways, dedicated lanes for autonomous trucks can be dynamically allocated. Major required changes in infrastructure for the deployment of CAVs are in the use of infrastructure, e.g., road certification and speed limits, as well as in design standard and guidelines, e.g., pavements, road alignment, road structures, cross-section, pavement loadings, intersections, highways.

We notice that some efforts have been done in unifying safety standards by relevant professional societies and associations, namely International Organization for Standardization (ISO), Society of Automotive Engineers (SAE) and Institute of Electrical and Electronics Engineers (IEEE). There are only three standards that are directly related to CAV-enabled infrastructure, including ISO 37181:2022, ISO 37168:2022 and ISO 22737:2021. Most other existing standards are related to CAV alone, or transportation infrastructure alone (Table 6). No relevant standards are found in respective national bodies such as American National Standards Institute (ANSI) despite CAVs have been tested and operated in existing infrastructure for long. 

\subsection{Coexistence of different types and penetration rates of CAVs complicating digital infrastructure supports}

The greater benefits will be achieved dedicated infrastructures are constructed for the CAVs. For example, CAV-dedicated pavements are reckoned to bring forward a safer and efficient transportation \cite{schieben2019designing,faisal2019}. The conflicts between different types of vehicles will be reduced for both high-speed and low-speed modes in the CAV-dedicated areas. The disadvantage is that such designated areas are expensive to build, which require strong government supports, take a long time to implement and may bring benefits mainly to CAV users.

In other words, the benefits of the infrastructure are not the same for different types of the CAVs. The CAVs at highest level can make use of small areas of the lanes by effective and efficient coordination through platoon optimization. The AV with an automation level 2 or 3 may have a lack of connection and thus is limited by its perception sensor capacity. As a result, it may require more longitudinal and lateral spaces in the road in responding to undesirable behavior, leading to a smaller number of spaces. More importantly, with the mixture of different types of CAVs, the benefits of traffic optimization are unlikely to be achieved. More importantly, there is a large uncertainty on the timing for which the CAV at level 5 to become available and widely accepted \citep{soteropoulos2020automated}.

On the other hands, the coexistence of different types of CAVs may require more complex infrastructure supports. For example, the AVs may rely on retro-reflective materials to facilitate the safe perception whilst they may ignore the signals from road-side units. At intersections, the AVs may require the existence of regular traffic lights and signs. In contrast, the CVs do not require or expect a minimal existence of retro-reflective materials. At intersection, digital communications are enough for the CVs to operate safely. A survey summarizing expert opinions suggested that by fully CAVs are not ready in coming decades but the ones with lower level will be commercialized instead  \citep{tabone2021vulnerable}. \cite{soteropoulos2020automated} highlighted the importance of research on how to determine an area that are suitable for a certain type of CAVs, especially the level 4 vehicles.

\cite{alessandrini2021role} suggested that there should be a standard to certify ready infrastructure as well as CAVs. \cite{soteropoulos2020automated} proposed a so-called automated drivability index (ADI) to measure the drivability of CAVs within the transport network. However, the elements of the standard and parties involved in certification have not been determined, yet to mention its technical specifications. When it is likely that CAVs are mostly electric, the locations and standardization of stations for batter charging and swapping should be part of the planning \citep{azin2021infrastructure, vosooghi2020shared}. 
6.4. The integration of CAVs transforming the infrastructures but varying greatly subject to adopted strategies

\subsection{Integration of CAVs transforming digital infrastructures but varying greatly subject to adopted strategies}

The integration of CAVs will transform the infrastructures, and thus change the way we live and work
\citep{renne2022creating, orfanou2022humanizing}
. The adoption of CAVs may require significant changes in parking designated areas including smaller spaces and less location restrictions
\citep{zakharenko2016self, schieben2019designing, kellett2019might, manivasakan2021infrastructure}. The demand for parking is expected to be reduced dramatically since the movement of CAVs can be optimized for pickup and drop-off with a high level of automation \cite{faisal2019,schieben2019designing, kellett2019might}. A new way of transportation that make use of shared cars, namely Personal Rapid Transit (PRT) is promising since it uses a collection of autonomous pods that grouping a small batch of passengers in a CAV in a dedicated guideway  \citep{mittelman2022techno}. \cite{ziakopoulos2021quantifying} promoted the transport optimization of autonomous urban shuttle service (AUSS) to coordinate the use of general CAVs. From the urban planning perspective, \cite{medina2018exploring} suggested that in case the city radii increase 33\%, the travel time will increase 20\% but could be compensated by the benefits of CAVs.

However, there are currently a debate on which are the best strategies. The approaches to embrace the infrastructure can be categorized into three types: conservative, moderate, and aggressive strategies \citep{manivasakan2021infrastructure}. The conservative strategy with mixed traffic strategy encompasses certain upgrades to the current infrastructure where it is shared by both conventional vehicles and CAVs. No major change in the infrastructure and the construction of CAV-dedicated areas are made. This strategy gives an improvement in the safety while the penetration rate of CAVs reaches 25\% and provides a less congested traffic while the rate is higher than 5\% \citep{cui2017stabilizing}. It will benefit the least as compared to other two strategies; but its advantage is the less friction during the implementation and transition, and hence it may be more favored by the public. In the moderate strategy with autonomous corridors, both CAV-dedicated road areas alongside the current infrastructure co-existed and supported. Such exclusive lanes allow for faster travel times, safer travels, and more efficiency in operation, yet to mention improved experiences for the CAV-users as compared to the first approach. It requires less costs and frictions to redevelop new infrastructure and areas compared to the aggressive strategy. Its benefits will be clear while the penetration rate is 40\%. Despite the overall transportation network may be impacted by such restricted lanes, it is the most resilient and practical way for the transition toward the aggressive strategy with fully CAV-dedicated areas. The aggressive strategy with fully separated areas will transform entire areas into the fully separated CAV-permitted regions, such as in the city centers. The safety, efficiency, and many other benefits of the CAVs can be maximized due to the ability to coordinate and optimize the traffic with supported infrastructure. Within such areas, new dedicated infrastructure is needed alongside with significant redesigns and upgrades to existing ones to enable this strategy. This approach is costly and accompanied by various technical challenges to be resolved. It may not be well received by the public, at least in a near future. \cite{overtoom2020assessing} suggested that the improvement in designing the dedicated lanes may be good for reducing the delays but not be helpful in decreasing the congestions, especially when the penetration rate of CAVS is smaller than 25\%. 

Interestingly, \cite{liao2021shared} showed that the electric CAVs can act as vehicle-to-grid (V2G) services to stable the energy demands. In a study with a mid-size city (Ann Arbor, Michigan, United States), they found that the integration of V2G service to the CAVs is 19.6\% more economically feasible as compared to the shared CAVs in 30-year timeframe with the revenue of 2722 per CAV per year. \cite{hannan2022vehicle} shared the same opinion that this utilization will benefit both electric CAV owners and the power grid.

\subsection{Safety, cybersecurity, validation, and privacy putting a pressure for digital infrastructure to support CAVs}

CAVs promise to provide various benefits to safety and traffic network efficiency. Having said that, these benefits of CAVs are not possible without a comprehensive support from the infrastructure. Currently, the CAVs have been tested in different ways, using computer simulators, test loop or real-world on-road tests. While road tests are of importance, it is unknown whether certain upgrades and changes infrastructure are able to support the safe operation of the CAVs. However, various technical challenges in the design of a practical, efficient, and secure communications are needed to be overcome \citep{he20206g}. A safety problem can be arisen from the lack or untimeliness of HD map updates. A possible way is to use crowdsourced approach that make use of sensors from the CAVs to provide details and updates from the environment \citep{tchuente2021providing}. 

Digitised transport infrastructure is contributed to improve the operation and control of transport network, providing shorter travel times, reduce vehicles on road, less congestions, lower number of crashes thanks to V2I, V2V and V2X communications. However, it will bring not only benefits but also the risks, such as cybersecurity issues, for the operation of CAVs. In the aspect of cybersecurity, many studies \citep{deng2021deep, nikitas2022deceitful} pointed out that CAVs are prone to different types of physical and digital attacks during operation on roads. For regular AVs, attackers can introduce physical attacks on sensors using different techniques. With jamming methods, they could introduce extensive light blinding attack to cameras \citep{petit2015remote}, trigger waves having same frequencies as of LiDAR \citep{shin2017illusion}, create vibrations to gyroscopic sensors \citep{son2015rocking} by installing physical devices in the infrastructure or hijacking existing road-side devices to start the attacking. With spoofing techniques, they could introduce ``ghost'' objects into the LiDAR vision \citep{petit2015remote}, changing the sensor visions with fake signals \citep{yan2016can} or modifying the existing signs \citep{nassi2019mobilbye}. For the digital attacks, they can introduce adversarial attacks to deep learning models to fool the perception of objects and signs or sending fake data when the CAVs communicating with the cloud services through road-side unit equipment in loading maps, updating software, or obtaining the operation instructions \citep{deng2021deep}. 

When it comes to platoon movement of the CAVs, it is unclear how the privacy of CAVs is protected when sending and receiving the signals. For example, when the leading car can send its captured images and cloud points to surround vehicles or to traffic control unit, the information can be compromised to extract private information. The data may contain the network address, locations, and quality of the CAVs that send and receive the signals. It is unknown to what extent the CAV’s users are able to control the shared data, or which sources of information that they can consider as reliable to be used for their decisions, which was pointed out that such information is also prone to security and safety issues \cite{khattak2022active}.  
As a result, new privacy-preserving schemes should be developed and integrated into the smart infrastructure \cite{sucasas2016}. Testing and validations, including standardised testbed facilities and scenarios, should be agilely implemented whenever new infrastructure and new CAVs are released.

\section{Concluding remarks}

The revolution brought forward by the CAVs is not only in the aspect of automated driving and cooperative mobility, but also in the according change that they make while in transforming the current infrastructure. Our main findings in this survey are as follows. 
First, the deployment of CAVs may introduce significant impacts to existing infrastructure; and in its turn, the infrastructure may be unsafe for the CAVs. On one hand, the deployment of CAVs will optimize the traffic flow with higher throughput and less congestions, reduce the number of fletch and vehicles and decrease the accidents. On the other hand, current physical infrastructure such as pavement, traffic signs and devices, parking facilities as well as road equipment and elements supporting vulnerable road users are yet to be ready for the safe operation of the CAVs.  Further assessment is needed, for example, how the pavement surface may be suffered from intense loading when the CAVs operate in platooning mode. The adoption of very high-resolution and real-time digital maps, faster and more reliable communications protocol in the digital infrastructure will bring great benefits to the operational infrastructure, including traffic controls and optimization as well as transforming the urban development. 

Second, to enable the CAVs, significant changes in physical, digital, and operational infrastructure are required. For physical infrastructure, we may need new type of pavements that are reinforced by stronger materials, have reshaped geometries, and contain enhanced markings with retroreflective materials alongside new generations of road signs and traffic signals. Innovative designs for parking and service stations, upgrades, and rebuilding of infrastructure to support the CAVs in communicating with diverse range of road users, especially at intersection, are needed. For a cooperative and safe driving experience, new digital infrastructure is expected to be implemented, which include, for instances, high-resolution map layers for transportation and traffic (of which some elements can be updated in real time), the integration of V2I and V2V devices into road infrastructure and the development of new cooperative driving applications and systems. New generations of intelligent transport systems and better urban planning should consider the CAVs at different stages of deployment.

Third, several challenges and opportunities in research have been identified and discussed. There is currently lack of governmental guidelines and laws on CAV-enable infrastructure. The standardization of CAV-enabled infrastructure is also nonuniform across country, regions, and sectors. The coexistence of different types and penetration of CAVs complicate the infrastructure supports. The integration of CAVs will transform the infrastructures but it is subject to adopted strategies. The verification, safety, cybersecurity, and privacy cause a great pressure for the infrastructure to support CAVs.

\section*{Acknowledgement}

This research work is sponsored by the SPARC Hub (https://sparchub.org.au) at Department of Civil Engineering, Monash University, funded by the Australian Research Council (ARC) Industrial Transformation Research Hub (ITRH) Scheme (Project ID: IH180100010).

\bibliographystyle{apalike}
\bibliography{references}

\end{document}